\documentclass[preprint]{aastex}

\usepackage{graphicx}
\usepackage{natbib}
\usepackage{pdflscape} 
\usepackage{amssymb}
\usepackage{ulem} 
\usepackage{color} 

\newcommand{\be}{\begin{equation}}
\newcommand{\ee}{\end{equation}}

\newcommand{\ba}{\begin{eqnarray}}
\newcommand{\ea}{\end{eqnarray}}
\newcommand{\om}{\omega}

\newcommand\eg{{\it{{e.g.,\ }}}}

\newcommand{\Bf}{{magnetic field}}
\newcommand{\Bfs}{{magnetic fields}}
\newcommand{\Ef}{{electric  field}}
\newcommand{\Efs}{{electric fields}}

\newcommand{\NS}{neutron star}
\newcommand{\NSs}{{neutron stars}}

\newcommand{\ms}{magnetosphere}
\newcommand{\mss}{magnetospheres}

\begin{document}

\title{Fast Radio Bursts' emission mechanism: implication from localization}

\author{Maxim Lyutikov,\\
Department of Physics and Astronomy, Purdue University, 
 525 Northwestern Avenue,
West Lafayette, IN
47907-2036 \\
and \\
Department of Physics and McGill Space Institute, McGill University, 3600 University Street, Montreal, Quebec H3A 2T8, Canada}

\begin{abstract}
We argue that  the localization of the Repeating FRB at $\sim 1$ Gpc   excludes  rotationally-powered type of  radio emission (\eg analogues of Crab's giant pulses  coming from very young  energetic pulsars) as the origin of FRBs.
 \end{abstract}
\maketitle 

\section{Introduction}
 Fast radio bursts 
  \citep[FRBs,][]{2007Sci...318..777L,2012MNRAS.425L..71K,2013Sci...341...53T,2014ApJ...797...70K,2014ApJ...790..101S,2016Natur.531..202S} 
are recently identified type of transient radio emission.
The recent localization of the host galaxy of the repeating Fast Radio Burst 121102 (the Repeating FRB below)  by \cite{2017arXiv170101098C} establishes sources of FRBs as cosmological (and not only extragalactic) objects. The host galaxy is located at  the distance $D \sim 1$ Gpc.

The power and the (upper limit on the intrinsic) duration of FRBs require \NS-like  energy densities of the \Bf\ in the emission region   \citep{2016MNRAS.462..941L}. In the case of FRB 121102,  the  instantaneous (isotropic-equivalent)  luminosity $L_{FRB}$   is  
\be
L_{FRB} =4\pi D^2  (\nu F_\nu) \approx   10^{42}  F_{1 \rm Jy} D_{\rm Gpc}^2 {\rm erg \, s}^{-1},
\label{LFRB}
\ee 
where $F_\nu$ flux in Jansky and $\nu$ is the observation frequency.
Taking the duration of the bursts $\tau \approx 1$ msec $\equiv \tau_{-3}$  as indication of the emission size,
the  equipartition \Bf\ energy density at the source  is 
\be
B_{eq}=   \sqrt {8 \pi  } {\sqrt{\nu F_\nu } D \over c^{3/2} \tau} = 3  \times 10^8 \tau_{-3} ^{-1} 
\, {\rm G}.
\label{B2}
\ee
(For a beamed motion of the emitting plasma a similar \Bf\ would be required to collimate the beam.)
As the estimate (\ref{B2}) is the lower estimate on the \Bf\ at the source,  this narrows down the location of  the emission region to \mss\ of \NSs.

The brightness temperature
\be
T_b \approx {  2 \pi  D^2  F_\nu \over \nu^2 \tau^2}  {\Delta \Omega \over 4\pi} \approx 5 \times 10^{35} \, {\rm K}
\label{Tb}
\ee
clearly implies  a coherent mechanism. 

 In addition,  the estimate of the wave intensity parameter 
\be
a = {  e E \over m_e c  \om} \approx 10^5 \gg 1
\label{a}
\ee
where $E = \sqrt{L_{FRB} / ( c^3 \tau^2)}$ is the typical \Ef\ 
in the wave at the emission site and $\om = 2 \pi \nu$ is the observed frequency  \citep{2014ApJ...785L..26L}, require presence of strong \Bf\ with $\om_B \gg \om$, so that the correct definition of the intensity parameter  (\ref{a})  is  \citep{2016MNRAS.462..941L}
\be
a_B = {  e E \over m_e c  \om_B} \approx  1
\label{a1}
\ee
where $\om_B = e B/(m_e c)$ is the corresponding cyclotron frequency.

Thus,  \mss\ of \NSs\ are the most promising  {\it
loci} of the  FRBs emission generation
\citep{2010vaoa.conf..129P,2015ApJ...807..179P,2016MNRAS.457..232C,2016MNRAS.462..941L}.
Identification of FRBs with \NS\ and the repetitiveness as evidence against catastrophic events
(collapse, coalescence, etc.), leave two possible types  of the production of radio emission: 
(i) hypothetical radio emission accompanying giant flares in magnetars 
\citep{lyutikovradiomagnetar,2010vaoa.conf..129P,2014MNRAS.442L...9L,2012MNRAS.425L..71K,2015ApJ...807..179P,2016ApJ...826..226K}; 
(ii) Giant pulses (GPs) analogues emitted by young pulsars 
\citep{1995ApJ...453..433L,2004ApJ...616..439S,2007A&A...470.1003P}, 
see discussions by \cite{2016MNRAS.462..941L,2016MNRAS.457..232C,2016MNRAS.458L..19C}.
 These two possibilities rely on different source of energies for FRBs: 
strong magnetic fields in case of magnetars and the  rotational energy 
in case of GPs.

 \section{Not Giant Pulses from young energetic pulsars}

 \cite{2016MNRAS.462..941L} argued that if the FRBs  are analogues of giant pulses (GPs) but coming from young (ages tens to hundreds years) pulsars with Crab-like \Bf, then the required  initial periods need to be in a few msec range -  a reasonable assumption for $D \leq$ few hundreds Mpc.
 Identification of the FRB host with a galaxy at $D=1$ Gpc makes this possibility unlikely, as  we discuss next.

 For Crab pulsar the peak GP fluxes $F_\nu$  exceed Mega-Jansky
\citep{2003Natur.422..141H,2007whsn.conf...68S}. The observed fluxes from the Repeating FRB were in the hundreds milliJy range. Thus, it is required that the intrinsic GP power  at the FRB source is 
\be
\frac{L_{FRB}}{L_{GP}} \approx 2.5 \times 10^5
\ee
Scaling the FRB power with the spin-down power \citep[we note that this is not the case 
for the bulk of the pulsar population][]{ATNF} then puts constraints on the \Bf\ and the spin period of the FRB source:
\be
 \left(\frac{B_{FRB}}{B_{Crab}}\right)\left(\frac{P_{FRB}}{P_{Crab}}\right)^{-2}
\approx 500
\ee
($B$ and $P$ are corresponding surface \Bfs\ and periods). 
Thus, a Crab-like pulsar needs to spin at $1.5$ msec, while a magnetar-like \NS\ with quantum \Bf\ on the surface needs to spin at $\sim$  5 msec. Though these are physically allowed values,  as we show below the corresponding spin-down time is very short, contradicting  the observed constancy of the Repeating FRB over the few years of observations.

One possible caveat is that the FRB source can have higher efficiency in converting spin-down power into radiation than Crab's GPs (for Crab the {\it instantaneous}  efficiency can reach $10^{-2}$). 
Parametrizing the observed flux as
\be
\nu F_\nu = {\eta} \frac{L_{sd}} {4 \pi D^2},
\ee
where $L_{sd}$ is the spin-down luminosity an $\eta\leq 1$ is the conversion efficiency.
The conversion efficiency   $\eta$ is smaller than unity since the energy associated with an FRB should originate in the \mss\ of NSs (and not, \eg in the crust - this  would involve much longer time scales, $\sim 100$ msec -  the shear time scale through the crust). Also, pulsar glitches do not produce any considerable perturbation to the \ms. There is no way to ``store'' more energy in the \ms.

The longest possible spin-down time  is then
\be
 \tau_{SD} = \eta {\pi  I_{NS} \over 2 D^2  \nu F_\nu  P_{min}^2} \approx 600 \,   \eta \,  {\rm yrs}
 \label{tauSD}
 \ee
 for $F_\nu =1$ Jy and the  minimal period of $P_{min}=1 $ msec. (For a given FRB luminosity $L_{FRB}$, scaled with spin-down power, the longest  spin-down time is for shortest periods and, correspondingly, smallest \Bfs).
 The longest possible spin-down time scale (\ref{tauSD}) is barely consistent with constant value of the properties of the Repeating  FRB  over the period of few years - that would require an unrealistically high conversion efficiency $
\eta \rightarrow 1$. 

Another constraint comes from the constant values of the dispersion measure (DM) over few years \citep{2017arXiv170101098C}.
For the Repeating FRB approximately half of the DM contribution  comes from the local plasma, $DM _{loc} \sim  300$. Since, as we argued above, the typical time scale of the source is very short, a SN remnant should still be present. 
If DM is associated with an expanding SN shell, then it should sharply decrease with time, $DM \propto t^{-2}$   \citep{2016MNRAS.462..941L,2016ApJ...824L..32P}.
Since no DM changes are seen, this further excludes young rotationally-powered pulsars as FRB sources.

\section{Discussion}

The location of the Repeating FRB at  $\sim$ 1 Gpc \citep[an order of magnitude further away than what was a fiducial model in ][]{2016MNRAS.462..941L}, combined with a very steady value of DM, virtually excludes FRBs as analogues of Crab giant pulses. The allowed parameter region is very narrow: only an extremely efficient conversion of the rotational energy into radio waves, $\eta \approx 1$, {\it combined} with millisecond period  at birth and  a very specific range of ages $100$ yrs $<t$ $<500$ yrs (the lower limit come from the requirement of nearly constant DM, while the upper limit comes from the spin-down age, Eq. (\ref{tauSD})) can account for fluxes, duration, distance to FRB and the constant value of DM.

The alternative model -- hypothetical  generation of high brightness coherent radio emission during the initial stages of magnetar flares \citep[\eg][]{2003MNRAS.346..540L} --  is only marginally  better suited to explain FRBs. Briefly, we expect that during  the initial explosive stages of  the current sheet formation in  magnetar \mss\   the inductive \Efs\ accelerate particles \citep[somewhat   similar to acceleration mechanism discussed by][]{2016arXiv160305731L},  producing distributions unstable to generation of coherent emission. The peak luminosity of the 
 flare from SGR $1806-20$ was $10^{47}$ erg s$^{-1}$ \cite{palmer}. Thus,  efficiency of $\sim 10^{-5}$ between radio and $X$-rays is sufficient to power an FRB. But the corresponding signal from a magnetar at 10 $kpc$ would produce an FRB in the Giga-Jansky range. This may contradict the non-detection of the SGR $1806-20$ flare by Parkes  
 \citep{2016ApJ...827...59T}. We encourage an observational campaign to  detect possible  radio burst contemporaneous with magnetar bursts and flares.
 Another  puzzling property of  FRBs within the frameworks of both models  is  a very high and constant in time  local DM.

 I would like to thank members of  McGill Space Institute   (in particular  Victoria Kaspi  and Shriharsh Tendulkar),  Ue-Li Pen  and Sergey Popov for  discussions.

 This work was supported by   NSF  grant AST-1306672 and DoE grant DE-SC0016369.

 \bibliographystyle{apj} 
  \bibliography{/Users/maxim/Home/Research/BibTex}   \end{document}